\documentclass[manuscript]{raa}
\usepackage{graphicx,times}             
\usepackage{natbib}
\usepackage{amsfonts}
\usepackage{txfonts}

\newcommand{\Rmnum}[1]{\expandafter\@slowromancap\romannumeral #1@}
\begin{document}

\title{Metal Abundance and Kinematical Properties of M81 Globular Cluster System}
\volnopage{Vol.0 (200x) No.0, 000--000}      
\setcounter{page}{1}           

\author{Jun Ma\inst{1,2}\mailto{}
\and Zhenyu Wu\inst{1}
\and Tianmeng Zhang\inst{1}
\and Song Wang\inst{1,3}
\and Zhou Fan\inst{1}
\and Jianghua Wu\inst{4}
\and Hu Zou\inst{1}
\and Cuihua Du\inst{5}
\and Xu Zhou\inst{1}
\and Qirong Yuan\inst{6}
}

\institute{$^{1}$National Astronomical Observatories, Chinese Academy of Sciences, Beijing, 100012, P. R. China\\
$^{2}$Key Laboratory of Optical Astronomy, National Astronomical Observatories, Chinese Academy of
Sciences, Beijing, 100012, China\\
$^{3}$Graduate University, Chinese Academy of Sciences,
Beijing, 100039, China\\
$^{4}$Department of Astronomy, Beijing Normal University, Beijing 100875, China\\
$^{5}$College of Physical Sciences, Graduate University of the Chinese Academy of Sciences, Beijing 100049, China\\
$^{6}$Department of Physics, Nanjing Normal University, WenYuan Road 1, Nanjing 210046, China\\
\email{majun@nao.cas.cn}}


\abstract{In this paper, we presented metal abundance properties of 144 M81 globular clusters. These globulars consist of the largest globular cluster sample in M81 till now. Our main results are: the distribution of metallicities are bimodal, with metallicity peaks at ${\rm [Fe/H]}\approx -1.51$ and $-0.58$, and the metal-poor globular clusters tend to be less spatially concentrated than the metal-rich ones; the metal-rich globular clusters in M81 do not demonstrate a centrally concentrated spatial distribution as the metal-rich ones in M31 do; like our Galaxy and M31, the globular clusters in M81 have a small radial metallicity gradient. These results are consistent with those obtained based on a small sample of M81 globular clusters. In addition, this paper showed that there is evidence that a strong rotation of the M81 globular cluster system around the minor axis exists, and that rotation is present in the metal-rich globular cluster subsample, while the metal-poor globular cluster subsample shows no evidence for rotation. The most significant difference between the rotation of the metal-rich and metal-poor globular clusters occurs at intermediate projected galactocentric radii. The results of this paper confirm the conclusion of Schroder et al. that M81's metal-rich globular clusters at intermediate projected radii were associated with a thick disk of M81.
\keywords{galaxies: individual (M81) -- galaxies: star clusters -- globular clusters: general}}
\authorrunning{Jun Ma}
\titlerunning{Globular Clusters in M81}
\maketitle
\authorrunning{Ma}

\section{Introduction}

An understanding of galaxy formation and evolution is one of the principal goals of modern astrophysics. As the oldest stellar systems in the universe, globular clusters (GCs) maintain a fossil record of the early history
of galaxies, so they are important tracers of galaxy formation and evolution. In addition, since GCs are very dense, gravitationally bound spherical systems
of several thousands to more than a million stars, they can be observed out to much greater distances than individual stars.

The metallicity (or color) distribution of GCs is of particular importance in deepening our knowledge of the dynamical and chemical evolution of the parent galaxies. For example, based on data from the {\sl Hubble Space Telescope} ({\sl HST}) archive, \citet{Gebhardt99}, \citet{Larsen01} and \citet{kw01} showed that many
large galaxies possess two or more subpopulations of GCs that have quite different chemical compositions \citep[see also][]{west04}. \citet{peng06} presented the color distributions of GC systems for 100 early-type galaxies from the Advanced Camera for Surveys (ACS) Virgo Cluster Survey, and found that, on average, galaxies at all luminosities appear to have bimodal or asymmetric GC color/metallicity distributions. The presence of color bimodality among these old GCs indicates that there
have been at least two major star-forming mechanisms in the (early) histories of massive galaxies \citep{west04,peng06,strader06}.

\citet{Fan08} presented a metallicity distribution of 148
Galactic GCs \citep[][2003 updated version]{harris96} that apparently shows two peaks (i.e., two distinct metal-poor and metal-rich GC populations). A double-Gaussian can best fit these two subpopulations, the mean metallicity values are $-1.620$ and $-0.608$ dex, respectively. Using the data for 431 GCs in M31, \citet{Lee08} studied the metallicity distribution, which is asymmetric, implying the possibility of bimodality. The bimodal test results in the two group: the metal-poor with a mean value of $-1.47$ and the metal-rich with a mean value of $-0.62$ \citep[see also][]{bh00,perrett02,Fan08}. \citet{ma05} presented a metallicity distribution of 94 M81 GCs being bimodal, with metallicity peaks at $-1.45$ and $-0.53$.

M81 is one of the nearest Sa/Sb-type spiral galaxies outside the Local Group, very similar to M31, and roughly as massive as the Milky Way. So, beyond the Local Group, it is a good candidate for reaching a detailed study of a spiral galaxy GC system for comparison to the Milky Way and M31 GC system. In the pioneering work of M81 GCs, \citet{bh91} derived spectroscopic metallicities for eight GCs in M81 and presented the mean of $\rm [Fe/H]=-1.46\pm0.31$ for them. Then, \citet{pbh95} obtained low signal-to-noise spectra of 82 GC candidates, 25 of which were confirmed as $bona$ $fide$ M81 GCs. These authors derived the mean metallicity to be $\rm
[Fe/H]=-1.48\pm0.19$ both from the weighted mean of the individual metallicities, and directly from the composite spectrum of these 25 GCs. To maximize the success rate of M81 GC candidate list for the ongoing spectroscopic observations, \citet{pr95} used an extensive database that included photometric, astrometric, and morphological information on 3774 objects covering over a $>50'$ diameter field centered on M81 to reveal 70 GC candidates.

\citet{sbkhp02} presented moderate-resolution spectroscopy for 16 GC candidates from the GC list in \citet{pr95}, and confirmed these 16 candidates as $bona$ $fide$ GCs. They derived metallicities for 15 of the 16 GCs. From their results, \citet{sbkhp02} concluded that the M81 GC system is very similar to the Milky Way and M31 systems, both chemically and kinematically.

With the superior resolution of the {\sl HST}, M81 is close enough for its clusters to be easily resolved on the basis of image structure \citep{cft01}. Thus, using the $B$, $V$, and $I$ bands of {\sl HST} Wide Field Planetary Camera 2 (WFPC2), \citet{cft01} imaged eight fields covering a total area of $\sim 40~\rm{arcmin}^{2}$, and detected 114 compact star clusters in M81, the 59 of which are GCs. Based on the estimated intrinsic colors, \citet{cwl04} found that the M81 GC system has an extended metallicity distribution, which argues the presence of both metal-rich and metal-poor GCs.

\citet{ma05,ma06,ma07} presented detailed studies on M81 GCs based on the spectral and multi-color observations such as: the distributions of intrinsic $B$ and $V$ colors and metallicities of 95 GCs; the spectral energy distributions of 42 GCs in 13 intermediate-band filters from 4000 to 10,000 \AA, using the CCD images of M81 observed as part of the Beijing-Arizona-Taiwan-Connecticut (BATC) multicolor survey of the sky; the spatial and metal abundance properties of 95 GCs.

\citet{N10b} presented a catalog of extended objects in and around M81 based on an {\sl HST} ACS $I$-band mosaic, and found 233 good GC candidates. \citet{NH10} obtained spectra of 74 GCs in M81 with Hectospec on the 6.5 m MMT on Mt. Hopkins in Arizona, and determined their metallicities. \citet{NH10} analyzed the kinematics and metallicity of the M81 GC system combined their results with 34 M81 GC velocities and 33 metallicities from the literature.

In this paper, we will analyze the metallicity of the M81 GC system based on a large sample of GCs, and analyze the kinematics of the M81 GC system in details. The outline of the paper is as follows. In \S~2, we provide the analysis of the metallicity, and we provide the analysis of the kinematics in \S~3. The summary is presented in \S~4.

\section{Metallicity Analysis}

\subsection{A sample of GCs}

\begin{figure*}
\resizebox{\hsize}{!}{\rotatebox{-90}{\includegraphics{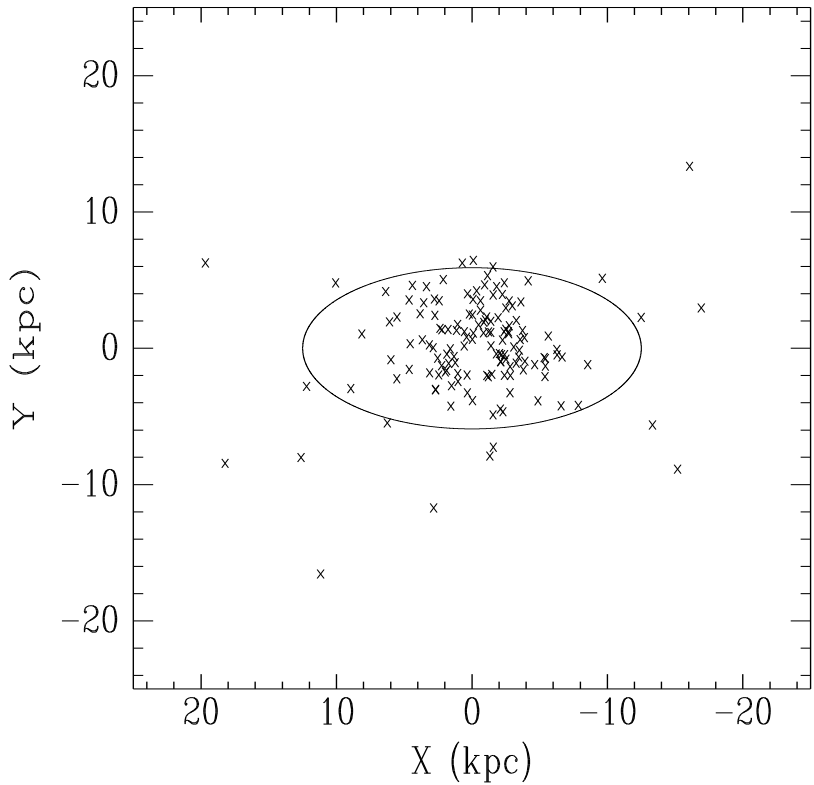}}} \vspace{0.0cm} \caption{Positions of the 144 M81 GCs. The large ellipse is the $D_{25}$ boundary of the M81 disk from \citet{Vaucouleurs91}.} \label{fig:one}
\end{figure*}

\citet{NH10} obtained spectra of 74 GCs in M81 with Hectospec on the 6.5 m MMT on Mt. Hopkins in Arizona, and determined their metallicities using a method based on the \citet{bh90} prescription, but with new index versus metallicity calibrations determined from the
41 Milky Way spectra from \citet{Schiavon05}, degraded to match the 5~\AA~resolution of their spectra \citep[see][for details]{N10a}. In order to increase the sample size and spatial extent, \citet{NH10} included the derived metallicities of 33 GCs in M81 from \citet{sbkhp02}, \citet{pbh95}, and \citet{bh91}. Altogether, there are 107 GCs in the \citet{NH10} sample of M81 GCs for performing metallicity analysis. \citet{ma05} studied the distributions of metallicities of 95 M81 GCs which are from \citet{pbh95}, \citet{cft01} and \citet{sbkhp02}. By cross-checking the sample GCs of between \citet{ma05} and \citet{NH10}, there are 51 GCs in common. We adopt \citet{NH10} metallicities with uncertainties smaller than 1.0 dex whenever available. However, when \citet{NH10} metallicity uncertainties are larger than 1.0 dex, we adopt the metallicities collected by \citet{ma05}. In addition, as \citet{ma05} pointed out that, since GC 96 of \citet{cft01} has very high $(B-V)_0$ ($(B-V)_0=1.778$), and the metallicity obtained using the color-metallicity correlation is too rich (0.95 dex), we do not include it in our sample of M81 GCs. Altogether, there are 144 M81 GCs in our sample.

Figure 1 shows the positions of 144 GCs in M81 with respect to the M81 disk. The $X$ coordinate is the position along the major axis of M81, where positive $X$ is in the northeastern direction, while the $Y$ coordinate is along the minor axis of the M81 disk, increasing towards the northwest. The relative coordinates of the M81 GCs are derived by assuming standard geometric parameters for M81. The distance modulus for M81 is adopted to be 27.8 \citep{fwm94,cft01}. We adopted a central position for M81 at $\rm \alpha_0=09^h55^m33^s.780$ and $\rm \delta_0=+69^o03'54''.34$ (J2000.0) following \citet{sbkhp02}. Formally,

\begin{equation}
X=A\sin\theta+B\cos\theta \quad ;
\end{equation}

\begin{equation}
Y=-A\cos\theta+B\sin\theta \quad ,
\end{equation}
where $A=\sin(\alpha-\alpha_0)\cos\delta$ and $B=\sin\delta \cos\delta_0 - \cos(\alpha-\alpha_0)
\cos\delta \sin\delta_0$. We adopt a position angle of $\theta=157^\circ$ for the major axis of
M81 \citep{cft01}.

\subsection{Metallicity distribution}

\begin{figure*}
\resizebox{\hsize}{!}{\rotatebox{-90}{\includegraphics{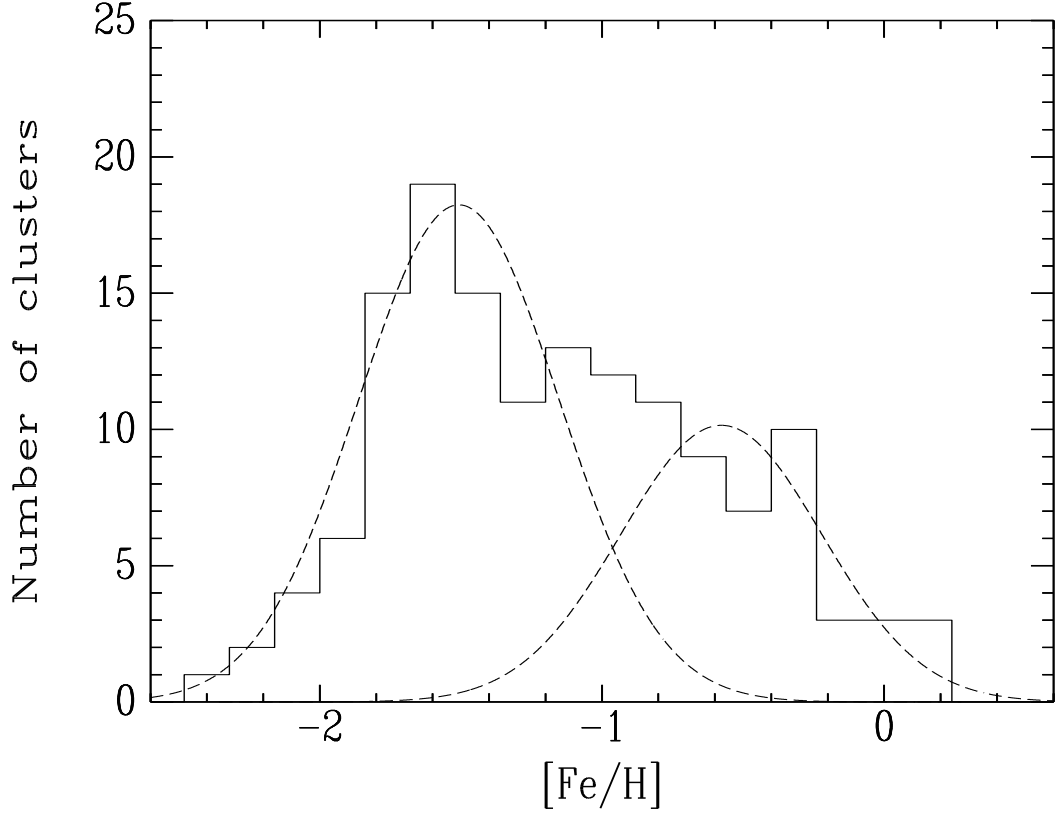}}} \vspace{0.0cm}
\caption{Histogram of the metallicities of 144 M81 GCs.} \label{fig:two}
\end{figure*}

Figure 2 displays the metallicity of 144 GCs in M81, and two metallicity peaks appear clearly. To make quantitative statements about the bimodal metallicity distribution, a KMM test \citep{abz94} is applied to the data. This test uses a maximum likelihood method to estimate the probability that the data distribution is better modeled as a sum of two Gaussians than as a single Gaussian.
Here we use a homoscedastic test (i.e., the two Gaussians are assumed to have the same dispersion). The metallicities of the two peaks, the $P$-value, and the numbers of GCs assigned to each peak by the KMM test are ${\rm [Fe/H]}\approx -1.51$ and $-0.58$, 0.02, and 91 and 53. The $P$-value is in fact the probability that the data are drawn from a single Gaussian distribution. It is evident that our results are in agreement with \citet{ma05} ($-1.45$ and $-0.53$) and \citet{NH10} ($-1.55$ and $-0.61$).

\subsection{Spatial distribution}

\begin{figure*}
\resizebox{\hsize}{!}{\rotatebox{-90}{\includegraphics{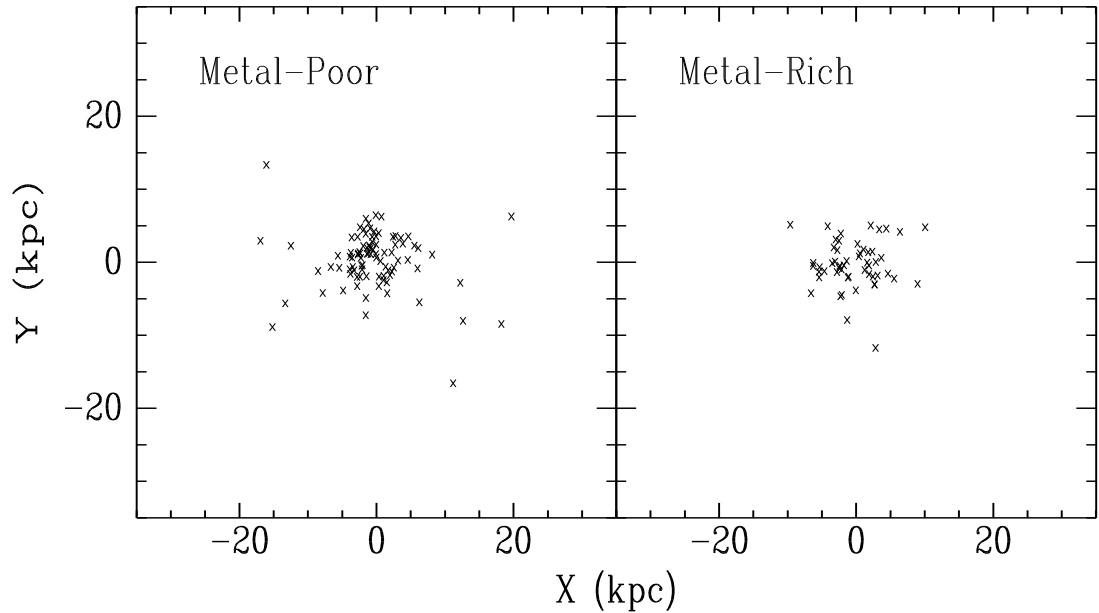}}} \vspace{-1.0cm} \caption{Spatial distributions of the metal-rich and metal-poor GCs in M81.} \label{fig:three}
\end{figure*}

Figure 3 shows the projected spatial distributions of the metal-poor and metal-rich GCs in M81. From Figure 3, it is clear that the metal-rich GCs in M81 are not as centrally concentrated as the metal-rich GCs of M31 are \citep{hbk91,perrett02,Fan08}, and the metal-poor GCs tend to be less spatially concentrated than the metal-poor GCs. These results confirm the conclusion of \citet{ma07} based on the small sample of M81 GCs.

\begin{figure*}
\resizebox{\hsize}{!}{\rotatebox{-90}{\includegraphics{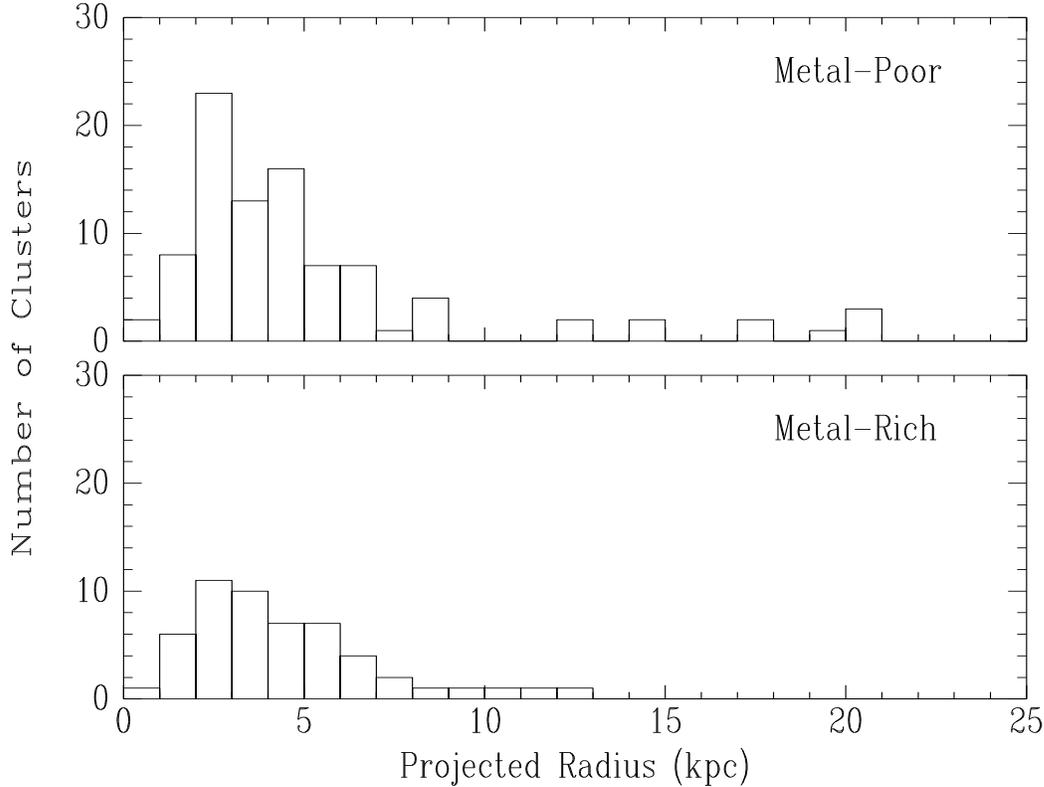}}}
\vspace{0.0cm} \caption{Radial distributions of the metal-rich and metal-poor GCs in M81.}
\label{fig:four}
\end{figure*}

\begin{figure*}
\resizebox{\hsize}{!}{\rotatebox{0}{\includegraphics{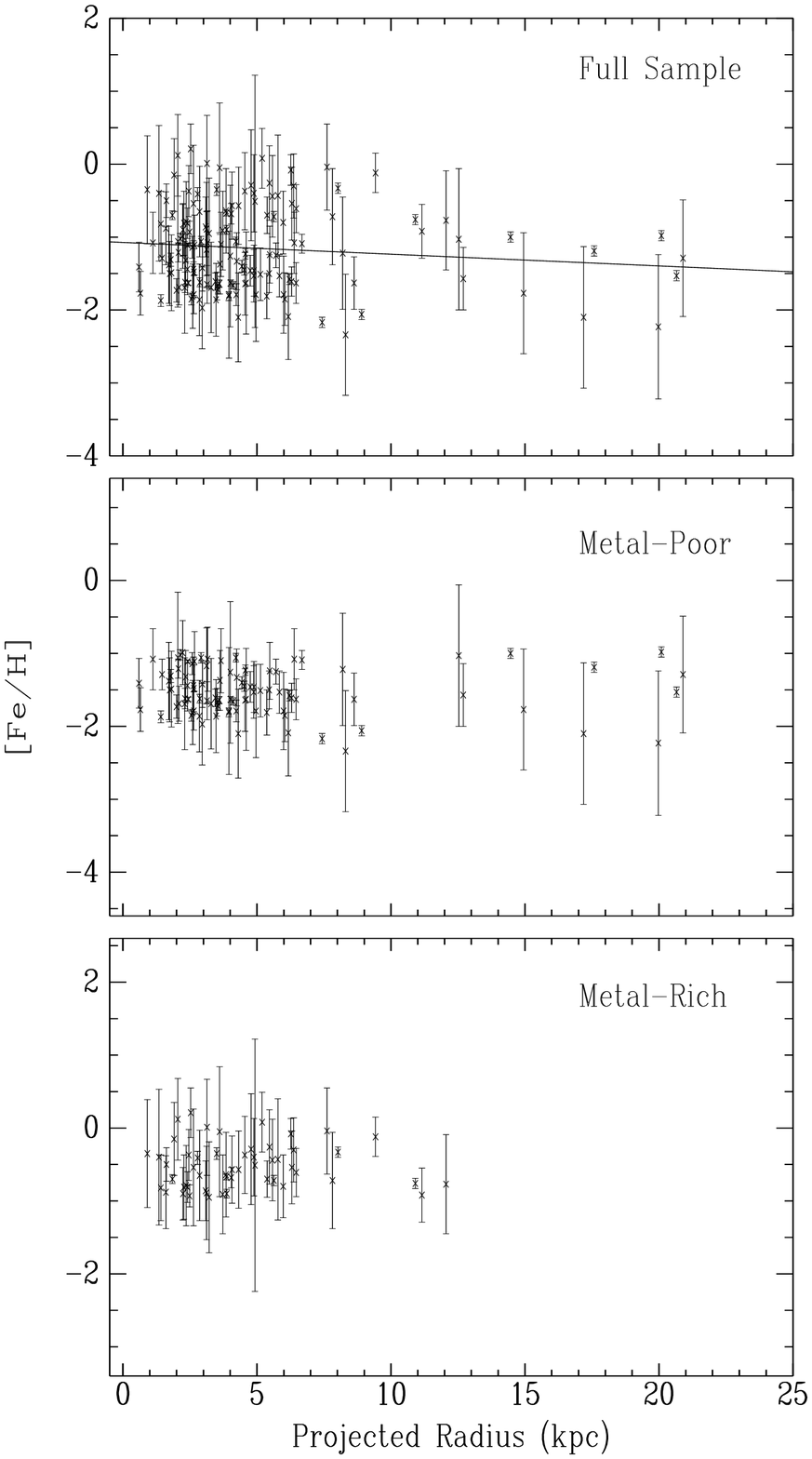}}}
\vspace{-1.5cm} \caption{Metallicity as a function of projected radius for M81 GCs.} \label{fig:five}
\end{figure*}

Figure 4 presents the histogram for the metal-poor and metal-rich GCs in M81. It shows that most of the metal-rich GCs distribute at projected radii of 1-7 kpc, and the peak lies at 2-4 kpc. It is also noted that the metal-rich GCs distribute within the inner 13 kpc, while the metal-poor GCs do out to radii of $\sim 20$ kpc. In the Milky Way, the metal-rich GCs reveal significant rotations and have historically been associated with the thick-disk system \citep{zinn85,taft89}; however, other works \citep{fw82,minniti95,patr99,forbes01} suggested that metal-rich GCs within $\sim 5$ kpc of the Galactic center are better associated with the bulge and bar. In M31, \citet{ew88} showed that the metal-rich GCs constitute a more highly flatted system than the metal-poor GCs, and appear to have disklike kinematics; \citet{hbk91} showed that the metal-rich GCs are preferentially close to the galaxy center. In addition, \citet{hbk91} showed that the distinction between the rotation of the metal-rich and metal-poor GCs is most apparent in the inner 2 kpc. So, \citet{hbk91} concluded that the rich-metal GCs in M31 appear to form a central rotating disk system. With a large sample of 321 velocities, \citet{perrett02} provided a comprehensive investigation on the kinematics of the M31 GC system. These authors showed that the metal-rich GCs of M31 appear to constitute a distinct kinematic subsystem that demonstrates a centrally spatial distribution with a high rotation amplitude, but do not appear significantly flattened, which is consistent with a bulge population. It is of interests to mention that, \citet{sbkhp02} performed a maximum-likelihood kinematic analysis on 166 M31 GCs of \citet{bh00} and found that the most significant difference between the rotation of the metal-rich and metal-poor GCs occurs at intermediate projected galactocentric radii. Especially, \citet{sbkhp02} presented a potential thick-disk population among M31's metal-rich GCs. For M81 GCs, \citet{sbkhp02} performed a kinematic analysis of the velocities of 44 M81 GCs, and strongly suggested that the metal-rich GCs are rotating in the same sense as the gas in the disk of M81. \citet{sbkhp02} concluded that, although their GC sample is not large enough to make a
direct comparison between metal-rich and metal-poor GCs in specific radius ranges, the conclusion with M81's metal-rich GCs at intermediate projected radii being associated with a thick disk in M81 is correct. In \citet{sbkhp02}, their sample is not large enough to make a direct comparison between metal-rich and metal-poor GCs in specific radius range. In this paper, with a larger GC sample, we can do this comparison (see \S 3). Our results will confirm the conclusion of \citet{sbkhp02} that M81's metal-rich GCs at intermediate projected radii is associated with a thick disk in M81.

\subsection{Metallicity gradient}

\begin{figure*}
\resizebox{\hsize}{!}{\rotatebox{-90}{\includegraphics{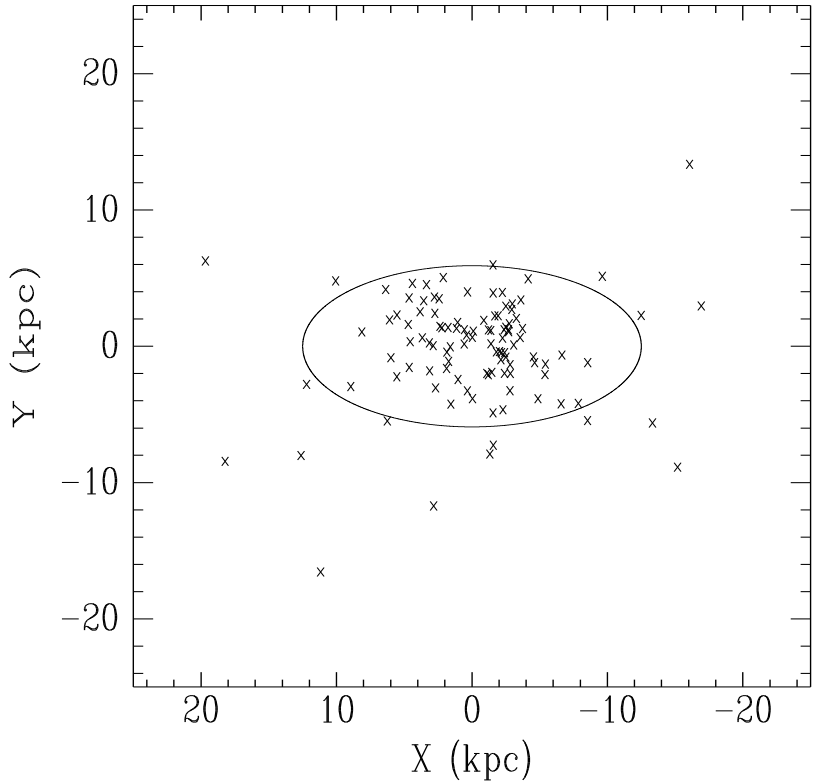}}}
\vspace{0.0cm} \caption{Positions of the 108 M81 GCs. The large ellipse is the $D_{25}$ boundary of the M81 disk from \citet{Vaucouleurs91}.} \label{fig:six}
\end{figure*}

One early formation model by \citet{eggen62} argued for a single, large-scale collapse of material to form galactic bodies such as the Milky Way, in which the enrichment
timescale is shorter than the collapse time, the halo stars and GCs should show large-scale metallicity gradients; however, \citet{sz78} presented a chaotic
scheme in the early evolution of a galaxy, in which loosely bound pre-enriched fragments merge with the main body of the proto-galaxy over a significant period, so there should be homogeneous metallicity distribution.

For the Milky Way, \citet{taft89} showed some evidence that metallicity gradients with both distance from the Galactic plane and distance from the Galactic center were present in the disk GC system. For M31, there are some inconsistent conclusions, such as \citet{van69} who showed that there is little or no evidence for a correlation between metallicity and projected radius, but most of his GCs were inside $50''$; however, some authors \citep[see, e.g.][]{hsv82,sha88,hbk91,bh00,perrett02} presented evidence for a weak but measurable metallicity gradient as a function of projected radius. \citet{Fan08} confirmed the latter conclusion based on their large sample of spectral metallicity and color-derived metallicity. Figure 5 plots the metallicity of the M81 GCs as a function of galactocentric radius. Clearly, the dominant feature of this diagram is the scatter in metallity at any radius. Since our sample GCs are mainly distributed in the inner 10 kpc, it is difficult to determine the metallicity gradient. It is true that, smooth, pressure-supported collapse models of galaxies are unlikely to produce a result like this. In order to present a quantitative conclusion, we made least-squares fits: the total sample of GCs does not have a significant metallicity gradient ($-0.016\pm0.012$ dex $\rm kpc^{-1}$), which confirm \citet{ma07} conclusion based on a smaller sample of M81 GCs. This result is also in agreement with \citet{kong00}, who obtained metallicity maps of M81 field by comparing simple stellar population synthesis models of BC96 \citep{bc96} with the integrated photometric measurements of the BATC photometric system, and did not find, within their errors, any obvious metallicity gradient from the central region to the bulge and disk of M81. In addition, we should emphasize that, in the least-squares fits of this paper, the metal-rich GCs seem to act an important part in determining the metallicity gradient.

\section{Kinematics}

\begin{figure*}
\resizebox{\hsize}{!}{\rotatebox{-90}{\includegraphics{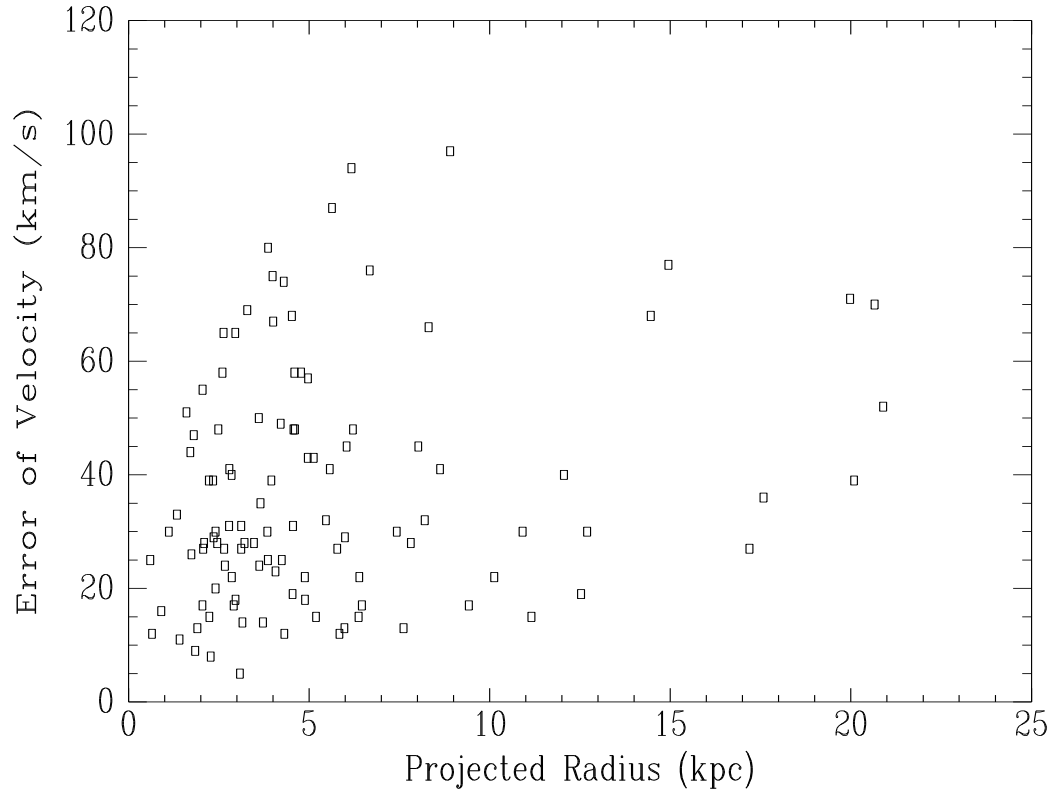}}}
\vspace{0.0cm} \caption{Errors of the measured velocities of M81 GCs versus a galactocentric radial distance.}
\label{fig:seven}
\end{figure*}

The M81 GC sample for studying kinematics is from \citet{NH10}, who obtained spectra of a total of
207 extended objects in M81 with Hectospec on the 6.5 m MMT on Mt. Hopkins in Arizona. These objects were selected from \citet{N10b}, who presented a catalog of extended objects in the vicinity of M81 based on a set of 24 {\sl HST} ACS
Wide Field Camera (WFC) F814W images, and found 233 good GC candidates. Based on these spectra, \citet{NH10} found 74 GCs in M81, 62 of which were newly confirmed by themselves. Combined with other 34 GCs from \citet{bh91}, \citet{pbh95}, and \citet{sbkhp02}, \citet{NH10} presented a catalog of radial velocities for 108 GCs in M81. Based on this catalog, \citet{NH10} studied the rotation, mean velocity, and velocity dispersion of M81 GCs, and presented a conclusion that, the M81 GC system as a whole shows strong evidence of rotation, with $V_r=108\pm22 {\rm km~s^{-1}}$ (a deprojected rotational velocity) overall. In this paper, we will present other evidence of rotation of the M81 GC system.

Figure 6 shows the positions of 108 GCs in M81 with respect to the M81 disk. Using the maximum likelihood method of \citet{pm93} \citep[see also][for details]{wu09}, we find a mean line-of-sight velocity for the full GC sample of $\langle v \rangle=-24\pm14~{\rm km~s^{-1}}$, somewhat higher than the systemic velocity of M31, $\langle v \rangle=-34\pm 4{\rm km~s^{-1}}$ (de Vaucouleurs et al. 1991). The GCs have an overall velocity dispersion of $\sigma_v=142\pm11~{\rm km~s^{-1}}$. In addition, the mean value of the radial velocities and the velocity dispersion are derived to be $\langle v \rangle=-22\pm25~{\rm km~s^{-1}}$ and $\sigma_v=153_{-16}^{+18}~{\rm km~s^{-1}}$ using the biweight location and scale of \citet{Beers90}. The kinematics of metal-rich and metal-poor GC populations will be discussed later in this section. First we examine the global kinematic properties of the M81 GCs. Figure 7 shows the errors of the measured velocities of M81 GCs versus a galactocentric radial distance. From Figure 7, we cannot find evidence that the mean errors for the measured velocities are larger in the inner region where background level is higher than that in the outer region of M81. Figure 8 displays the radial velocity histogram for 108 GCs. It appears that the velocity distribution for all GCs is roughly symmetric with respect to the mean value of the radial velocities

\begin{figure*}
\resizebox{\hsize}{!}{\rotatebox{-90}{\includegraphics{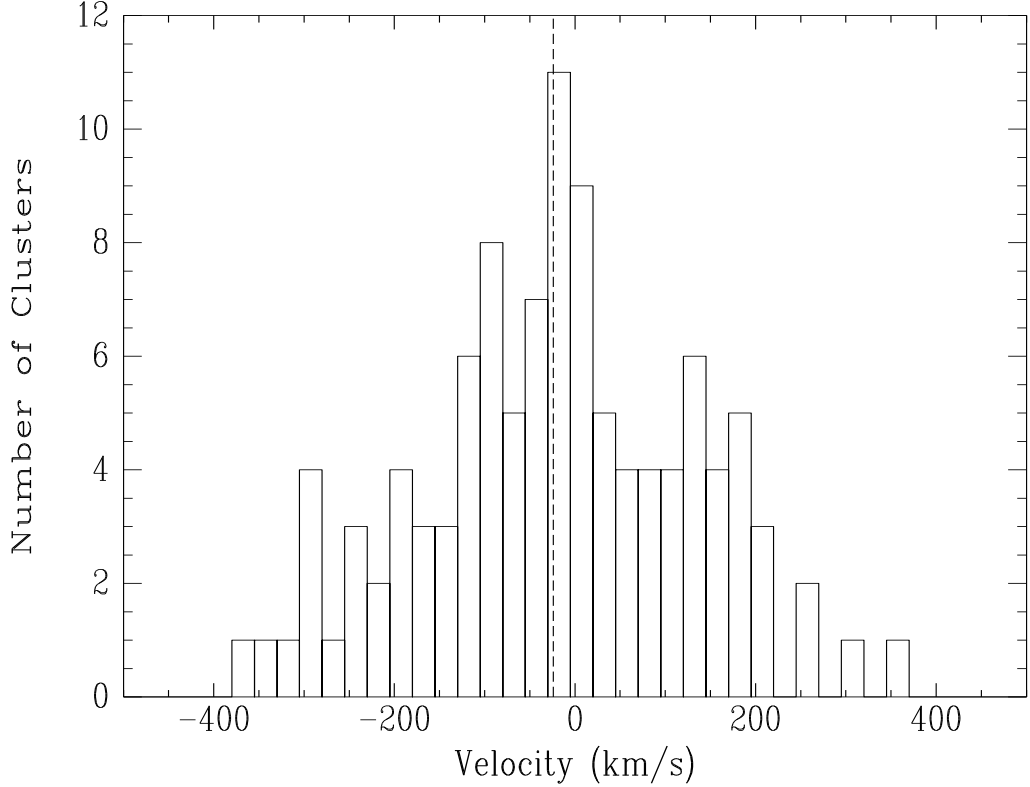}}}
\vspace{0.0cm} \caption{Radial velocity histogram for 108 GCs in M81. The vertical dashed lines indicates the mean value of the radial velocities, $\langle v \rangle=-24~{\rm km~s^{-1}}$.}
\label{fig:eight}
\end{figure*}

Figure 9 presents the spatial distribution of M81 GCs with the measured velocity. Here we adopted as the systematic velocity of M31 the value derived using the maximum likelihood method of \citet{pm93},
$\langle v \rangle=-24\pm14~{\rm km~s^{-1}}$. In Figure 9, we present the radial velocities for M81 GCs versus the projected radii along the major axis ($X$) and minor axis ($Y$), respectively. A linear least-squares fit, passing through ($X, v-\langle v \rangle$)$=$($0,0$), along the major axis results in $v-\langle v \rangle =(-11.8\pm 2.7)~X~{\rm km~s^{-1}}$, and passing through ($Y, v-\langle v \rangle$)$=$($0,0$), along the minor axis results in $v-\langle v \rangle =(-0.38\pm 5.0)~Y~{\rm km~s^{-1}}$, where $X$ and $Y$ are units of kpc. The large value of the slope between $v-\langle v \rangle$ and $X$ indicates a strong rotation of the M81 GC system around the minor axis, while the small value of the slope
between $v-\langle v \rangle$ and $Y$ indicates no significant rotation around the major axis.

\begin{figure*}
\resizebox{\hsize}{!}{\rotatebox{-90}{\includegraphics{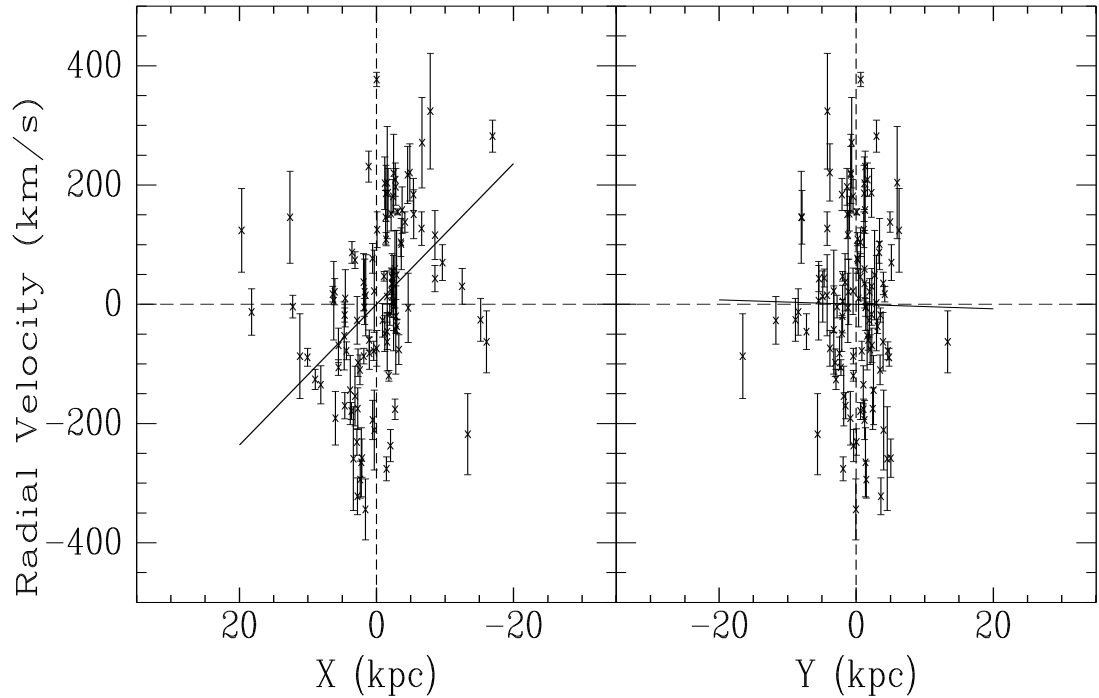}}}
\vspace{-0.5cm} \caption{Radial velocity of the M81 GCs versus projected radius along the major axis ({\it left}) and the minor axis ({\it right}). Solid lines indicate the linear least-squares fits.} \label{fig:nine}
\end{figure*}

Below, we will look for evidence for differences in subpopulations in M81 GCs that differ in their spatial and chemical characteristics. The formula for the rotation curve is $v=v_{\rm sys}+v_{\rm rot}\sin(\theta+\theta_0)$, where $v_{\rm sys}$ is the mean velocity of the GC system. The results of our kinematics analysis are summarized in Table 1. Figure 10 illustrates that rotation is present in the metal-rich GC subsample, while the metal-poor GC subsample shows no evidence for rotation. Table 1 also shows that the most significant difference between the rotation of the metal-rich and metal-poor GCs occurs at intermediate projected galactocentric radii. In addition, most of the metal-poor GC subsample have poorly constrained rotation axes and have rotation velocity-to-velocity dispersion ratios that are consistent with their being supported primarily by thermal motion rather than by rotation. These results are consistent with the conclusions of \citet{sbkhp02} for M31 GCs.

\begin{figure*}
\resizebox{\hsize}{!}{\rotatebox{-90}{\includegraphics{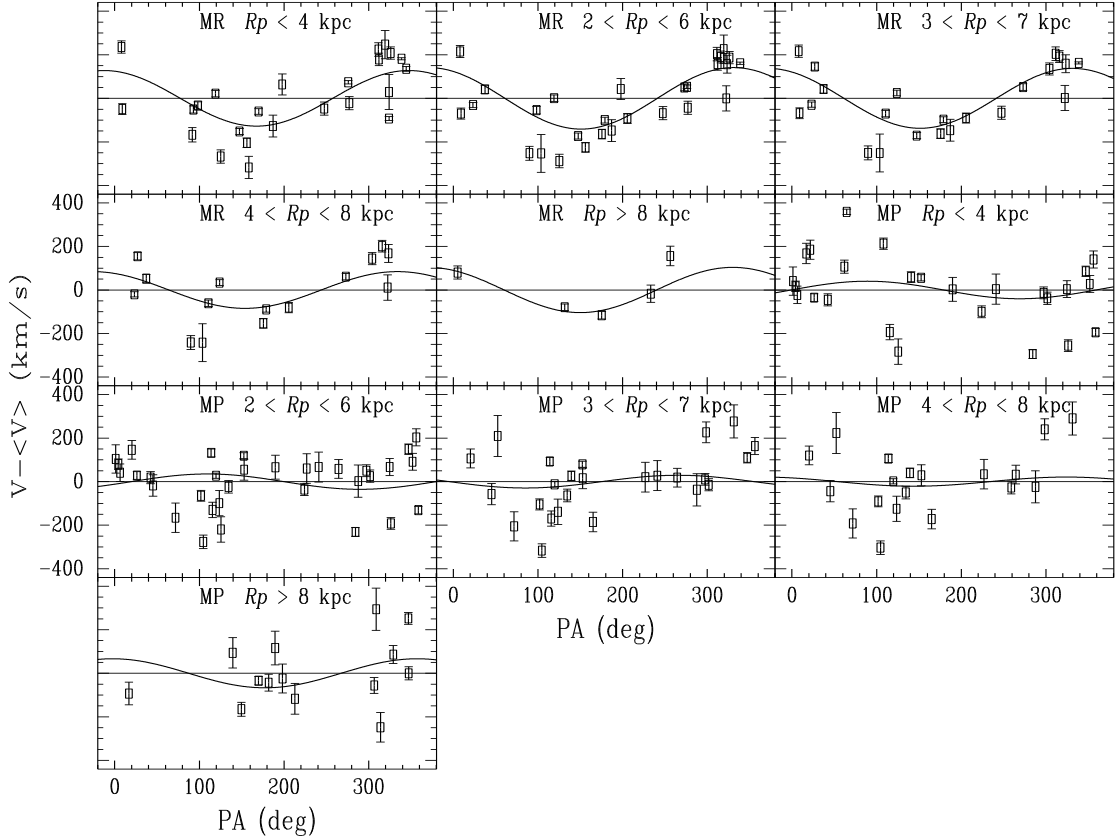}}}
\vspace{0.0cm} \caption{Quantitative assessment of rotation in M81 metal-rich (MR) and metal-poor (MP) GC subsamples based on projected galactocentric radius ($Rp$).} \label{fig:ten}
\end{figure*}

\section{Summary}

In this paper, we presented metal abundance properties of
144 M81 GCs which consist of the largest GC sample in M81. We also studied the kinematics properties of the M81 GC system. Our main conclusions are as follows:

1. The distribution of metallicities are bimodal, with metallicity peaks at ${\rm [Fe/H]}\approx -1.51$ and $-0.58$.

2. The metal-poor GCs tend to be less spatially concentrated than the metal-rich ones.

3. The metal-rich GCs in M81 do not demonstrate a centrally concentrated spatial distribution as the M31 metal-rich GCs do.

4. The GCs in M81 have a small radial metallicity gradient.

5. There is evidence that a strong rotation of the
M81 GC system around the minor axis exists.

6. Rotation is present in the metal-rich GC subsample, while the metal-poor GC subsample shows no evidence for rotation, and the most significant difference between the rotation of the metal-rich and metal-poor GCs occurs at intermediate projected galactocentric radii.

7. M81's metal-rich GCs at intermediate projected radii are associated with a thick disk in M81. This result confirms the conclusion of \citet{sbkhp02}.

\section*{Acknowledgments} 
We would like to thank the anonymous referee for providing rapid and thoughtful report that helped improve the original manuscript greatly. This work was supported by the Chinese National Natural Science Foundation grands No. 10873016, and 10633020,  10803007, 11003021, 11173016 and 11073032, and by National Basic Research Program of China (973 Program), No. 2007CB815403.

\begin{table}[ht]
\centering \caption[]{Kinematics of M81 globular clusters.}
\begin{tabular}{@{}cccccc@{}}
\hline
 Spatial Group  &  $v_{\rm sys}$    & $\sigma_{v_{\rm sys}}$ &   $v_{\rm rot}$ & $\theta_{0}$ & number \\
                & (km/s)            & (km/s)                 & (km/s)          &   (deg)            &        \\
\hline
 $R_p < $ 4kpc  &                   &                        &                 &            &        \\
 MR.......... & $-50.85\pm31.76$ & $152.07\pm22.80$ & $128\pm27$ & $102.50\pm19.05$ &   24 \\
 MP.......... & $-5.85\pm30.19$  & $149.64\pm21.51$ & $40\pm56$  & $0.92\pm69.90$   &   26 \\
 2kpc $<R_p<$ 6kpc  &                   &                        &                  &        &            \\
 MR.......... & $-39.02\pm28.72$ & $148.32\pm28.72$ & $141\pm15$ & $119.00\pm18.16$ &   28 \\
 MP.......... & $-69.12\pm20.69$ & $111.89\pm14.95$ & $35\pm28$  & $-8.20\pm54.86$  &   33 \\
 3kpc $<R_p<$ 7kpc  &                   &                        &                  &        &            \\
 MR.......... & $-31.77\pm30.23$ & $138.27\pm22.37$ & $138\pm18$ & $118.60\pm13.06$ &   22 \\
 MP.......... & $-29.52\pm27.08$ & $127.30\pm20.81$ & $29\pm40$  & $184.78\pm163.35$&   25 \\
 4kpc $<R_p<$ 8kpc  &                   &                        &                  &        &            \\
 MR.......... & $-40.97\pm33.98$ & $127.58\pm25.49$ & $84\pm21$  & $116.01\pm43.54$ &   15 \\
 MP.......... & $-43.15\pm32.67$ & $133.62\pm25.43$ & $21\pm39$  & $125.99\pm340.47$&   19 \\
 $R_p>$ 8kpc    &                   &                        &                 &        &            \\
 MR.......... & $-34.03\pm48.97$ & $104.72\pm42.02$ & $104\pm27$ & $119.92\pm51.11$ &   5  \\
 MP.......... & $6.06\pm36.57  $ & $125.83\pm28.42$ & $67\pm42$  & $92.59\pm90.44$  &   14 \\
\hline
\end{tabular}
\end{table}


\end{document}